\documentclass[12pt]{iopart}

\usepackage{iopams}  

\expandafter\let\csname equation*\endcsname\relax

\expandafter\let\csname endequation*\endcsname\relax
\usepackage{amsmath}

\usepackage{graphicx}
\usepackage{enumerate}

\begin{document}

\title{ Nexus  and Dirac lines in topological materials}

\author{T.T. Heikkil{\"a}$^1$\footnote{Tero.T.Heikkila@jyu.fi} and
  G.E. Volovik$^{2,3}$\footnote{volovik@boojum.hut.fi}}
\address{$^1$ University of Jyvaskyla, Department of Physics and
  Nanoscience Center, P.O. Box 35, FI-40014 University of
  Jyv\"askyl\"a, Finland}

\address{$^2$ Low
Temperature Laboratory, Aalto University, P.O. Box 15100, 00076
Aalto, Finland} \address{$^3$ L.~D.~Landau Institute for
Theoretical Physics, 117940 Moscow, Russia}

\date{\today}

\begin{abstract}
We consider the $Z_2$ topology of the Dirac lines, i.e., lines of band
contacts, on an example of graphite. Four lines --- three with
topological charge $N_1=1$ each and one with $N_1=-1$ --- merge
together near the H-point and annihilate due to summation law
$1+1+1-1=0$. The merging point is similar to the real-space nexus, an
analog of the Dirac monopole at which the $Z_2$ strings terminate.
\end{abstract}

\vspace{2pc}

\maketitle


\section{Introduction}

Dirac points in 2D systems and Dirac lines in 3D are examples of the
exceptional points and lines of level crossing  analyzed by von
Neumann and Wigner \cite{Neumann1929}. They are typically protected by symmetry and are described by the $Z_2$ topological invariant (see Ref.~\cite{Horava2005}). Close to the 2D Dirac point 
with nontrivial topological charge $N_1$, the energy spectrum after deformation can be represented by the $2\times 2$ matrix 
$H_{\rm Dirac}=c(\sigma_1 p_x \pm \sigma_2 p_y)$, where the Pauli
matrices $\sigma_{1,2}$ describe the  pseudo-spin induced in the vicinity of the level crossing. This gives rise to the conical spectrum near the Dirac point 
$E=E_0 \pm c|{\bf p}_\perp|$.
For Dirac lines in 3D, the components $p_x$ and $p_y$ are in the transverse plane.

 The Dirac Hamiltonian anticommutes with $\sigma_3$, which allows us to have the analytic form for 
 the  topological charge $N_1$, see e.g. review \cite{Volovik2011}:
 \begin{equation}
N_1=  {\bf tr} \oint_C \frac{dl}{4\pi i} \cdot[\sigma_3 H_{\rm Dirac}^{-1}({\bf p}) \partial_l H_{\rm Dirac}({\bf p})]\,.
\label{eq:N1}
\end{equation}
Here  $C$ is an infinitesimal contour in momentum space around the Dirac point or the Dirac line.
The  topological charge $N_1$ in Eq.~(\ref{eq:N1}) is integer: $N_1=1$ for sign $+$ and $N_1=-1$ for sign $-$. However, the integer-valuedness emerges only in the vicinity of the Dirac point or line. In general, the summation rule is $1+1=2\equiv 0$. This means that the Dirac line with $N_1=2$ can be continuously deformed to the trivial configuration.

In time reversal symmetric superconductors, due to chiral symmetry the Dirac lines may have zero energy and thus correspond to the nodal lines in the spectrum. Such lines exist in cuprate superconductors. According to the bulk-boundary correspondence the nodal lines may produce a dispersionless spectrum  on the boundary -- the flat band with zero energy.
\cite{Ryu2002,SchnyderRyu2011,SchnyderBrydon2015} If time reversal
symmetry is violated (for example, by supercurrent), the Dirac line
acquires a nonzero energy, a Fermi surface is formed, and the Dirac line
lives inside the Fermi surface. In cuprate superconductors, the Fermi
surfaces created by supercurrent around Abrikosov vortices give rise,
at zero temperature, to the finite density of states proportional to
$\sqrt{B}$, where $B$ is the magnetic field. \cite{Volovik1993} 

Examples of semimetals with 2D Dirac points and 3D Dirac lines are provided by graphene and graphite, correspondingly.
In these materials the spin-orbit interaction can be ignored, and one can consider them as spinless materials. In graphene there are two Dirac points in the Brillouin zone. If graphene is treated as spinless, one Dirac point has
$N_1=+1$  and another one has $N_1=-1$ (for spinful electrons these are $N_1=+2$  and  $N_1=-2$ correspondingly). 

In rhombohedral graphite (ABCABC... stacking) two Dirac points of graphene layers generate
two well-separated Dirac lines.  Each has a form of a spiral and is characterized by the topological charge $N_1=+1$  or $N_1=-1$. \cite{HeikkilaVolovik2011,HeikkilaKopninVolovik2011} 
These Dirac lines have nonzero energy and
live inside the chains of the hole and electron Fermi surfaces, discussed by McClure (see Fig.~2 in Ref.~\cite{mcclure57}). The projection of the spiral to the top or bottom surface of graphite determines the boundary of the (approximate) surface flat band. \cite{HeikkilaVolovik2011,HeikkilaKopninVolovik2011} 
This flattening of the spectrum has been recently observed in epitaxial rhombohedral multilayer graphene.\cite{Pierucci2015}

\section{Bilayer graphene}

\begin{figure}
\centerline{\includegraphics[width=1.0\linewidth]{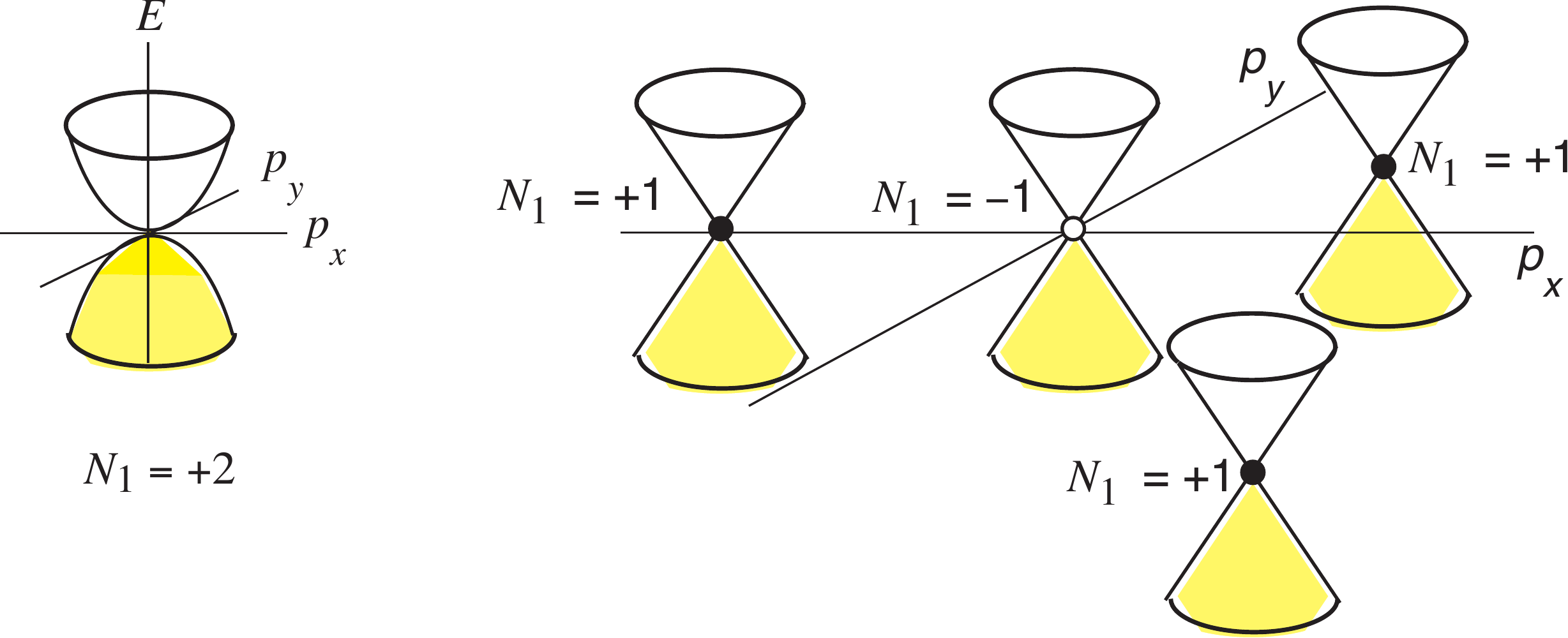}}
\caption{ \label{Fig:BilayerGraphene}
 $Z_2$-topology of Dirac points in bilayer graphene. {\it left}: Dirac
 point with quadratic spectrum characterized by topological invariant
 $N_1=2$. Since the topological invariant is trivial in $Z_2$
 topology, such a spectrum is unstable towards the gap formation, or to splitting. 
 {\it right}: Splitting to topologically stable Dirac points due to trigonal warping.
}
\end{figure}

In Bernal graphite (ABAB... stacking) the geometry of the Dirac
lines is essentially different from rhombohedral graphite, because  
Bernal graphite is the 3D extension of bilayer graphene with AB stacking. 
Let us hence start by describing the  bilayer graphene with AB stacking.

Since in each graphene layer  the topological charges are $N_1=+1$
and $N_1=-1$ in the two valleys, in bilayer graphene
the charges are summed up giving rise to trivial topological charges $N_1=+2\equiv 0$  and $N_1=-2\equiv 0$. Interaction between the layers may lead to several possible scenarios of the geometry of the fermionic spectrum in bilayer graphene:
\begin{enumerate}[(i)]
\item 
  If there is some special symmetry, such as the fundamental Lorentz invariance, one obtains a doubly degenerate conical spectrum.
However, in bilayer graphene there is no symmetry which could support
such a scenario.
\item The  topological charge $N_1=2$ gives rise to the Dirac fermions
  with parabolic energy spectrum near the Dirac point,  $E=E_0 \pm
  {\bf p}^2_\perp/2m$ (Fig. \ref{Fig:BilayerGraphene} {\it
    left}). Such a spectrum belongs to the trivial element of the group $Z_2$, that is why the neglected hopping elements destroy the parabolic spectrum. The Dirac point disappears, and finite gap (the Dirac mass) emerges.
\item The Dirac point with $N_1=+2$ splits into two Dirac points, each with $N_1=+1$ and with conical spectrum
(see \cite{KlinkhamerVolovik2005a} for the relativistic 3+1 system).
Such splitting, however, violates the hexagonal symmetry of graphene.
\item The splitting of the  $N_1=+2$ Dirac point, which is consistent  with  the hexagonal symmetry,  is in Fig. \ref{Fig:BilayerGraphene} {\it right}. The $N_1=+2$ Dirac point splits  into four Dirac conical points: 
 the central Dirac point has $N_1=-1$, while three others with $N_1=+1$ each are connected by the $C_3$ symmetry, see Fig. \ref{Fig:BilayerGraphene} {\it right}.
This is called the trigonal warping. 
The topological charges $N_1=\pm 1$ belong to the nontrivial element
of the group $Z_2$. That is why each Dirac point is topologically
stable and is not destroyed by addition of the neglected hopping
elements, if they obey the time reversal and sublattice symmetry. 
\end{enumerate}
The bilayer graphene chooses the scenario (iv) with the trigonal 
warping, \cite{CannFalko2006,KoshinoAndo2006} 
which prohibits the annihilation of the Dirac point due to splitting into the topologically stable Dirac points with $N_1=\pm 1$. Altogether the Brillouin zone of the bilayer graphene contains 8 Dirac conical points.

\section{From topology of bilayer graphene to Bernal graphite}

The trigonal warping also stabilizes  the Dirac lines in the Bernal graphite (ABAB... stacking).
Altogether the Brillouin zone of the Bernal graphite contains 8 Dirac lines,\cite{Mikitik2006,Mikitik2008}
which originate from 8 Dirac conical points of  bilayer graphene. These Dirac lines have finite energy and live inside the Fermi surface of graphite.

\begin{figure}
\centerline{\includegraphics[width=0.5\linewidth]{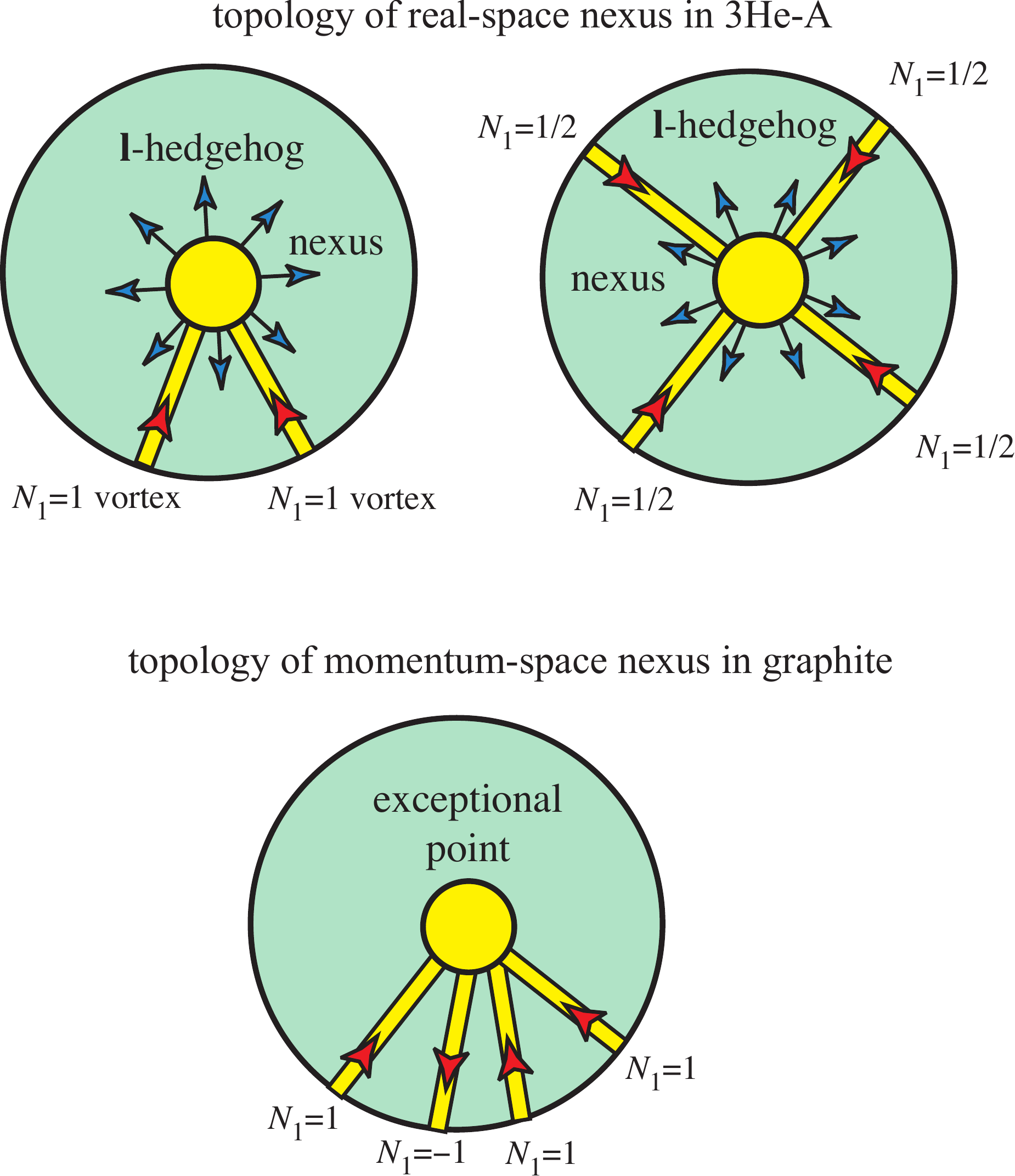}}
\caption{ \label{Fig:Nexus}
Real-space and momentum space nexus.
{\it top left}: Real-space nexus in the dipole locked superfluid $^3$He-A, where vortices are described by $Z_2$-topology. Two vortices, with $N_1=1$ each, merge together at the hedgehog in the field of the orbital momentum vector $\hat{\bf l}$ and annihilate each other since $N_1=2\equiv 0$.
{\it top right}: Real-space nexus in the dipole unlocked superfluid $^3$He-A, where vortices are described by $Z_4$-topology. Four half-quantum vortices, with $N_1=1/2$ each, merge together at the hedgehog and annihilate each other.
{\it bottom}: Illustration of the topology of the momentum-space nexus
in graphite. Dirac lines obey the $Z_2$-topology. Four Dirac lines,
with $N_1=+1$,   $N_1=+1$,  $N_1=+1$ and  $N_1=-1$, merge together at
an exceptional point of the spectrum and annihilate each other, because $N_1=1+1+1-1=2\equiv 0$.
}
\end{figure}

However, the topologically stable Dirac lines appear only in a certain region of momenta $p_z$. Outside that region the scenario (ii) without the Dirac lines takes place.
This means that there exists an exceptional point in the spectrum, at which four topologically stable Dirac lines merge together and annihilate each other. \cite{Mikitik2006,Mikitik2008}
Such a point is a direct analog of the nexus, a point in the real space where the $Z_2$ or $Z_4$ topological defects merge and annihilate each other, see Fig.~\ref{Fig:Nexus} {\it top}. \cite{volovikbook}

Let us find this exceptional point. According to  $D_{6h}$ symmetry of graphite, the Dirac lines 
are situated in the vertical mirror planes, and thus the exceptional point is on the H-K-H line in  the Brillouin zone at $p_z=P^*$.

The Hamiltonian of Bernal graphite is a 3D extension of the 2D Hamiltonian describing bilayer graphene:
\begin{equation}
H= 
\begin{pmatrix} 
 \Delta &v_F pe^{-i\phi}& \gamma_4  \Gamma(p_z) v_F pe^{i\phi}&  -2\gamma_1\Gamma(p_z) 
\\ 
v_F pe^{i\phi} & \gamma_2(p_z)& 2\gamma_3 \Gamma(p_z) v_F pe^{-i\phi}
&\gamma_4\Gamma(p_z) v_F pe^{i\phi}
\\
 \gamma_4 \Gamma(p_z) v_F pe^{-i\phi}&  2\gamma_3  \Gamma(p_z)
 v_F pe^{i\phi}  &\gamma_2(p_z) & v_F pe^{-i\phi}
 \\
-2\gamma_1\Gamma(p_z)  &  \gamma_4  \Gamma(p_z) v_F pe^{-i\phi} &v_F pe^{i\phi} & \Delta 
\end{pmatrix} \,.
\label{eq:HB}
\end{equation}
This corresponds to a rotated version of that written in
Ref.~\cite{mcclure57} (we write the Hamiltonian in a basis
spanned by the two layers $\times$ sublattice points, whereas
Ref.~\cite{mcclure57} writes it in the eigenbasis defined by the
strongest interlayer couplings), but we assume that the intralayer
coupling $\gamma_0$ dominates over the other terms, and therefore also
make the
$\sigma \cdot p$ approximation for each layer. Here $pe^{i\phi}=p_x + ip_y$; $\Gamma(p_z)=
\cos ( (\pi/2a)p_z)$, where $a$ is the distance between the K-point
and H-point in the Brillouin zone of graphite and  $p_z=0$ at the K-point
so that $\Gamma(K)=1$ and $\Gamma(H)=0$. The function
$\gamma_2(p_z)=\gamma_2 \Gamma^2(p_z)/2$ describes the coupling across
two layers  and $\Delta$ denotes the locally broken A-B sublattice
symmetry, which still preserves the global A-B symmetry. The
coefficients $\gamma_{3,4}$ are related to the original tight-binding
coefficients \cite{mcclure57} $\tilde \gamma_{3,4}$ via $\gamma_{3,4}
= \tilde \gamma_{3,4}/\gamma_0$. In graphite, $\gamma_3 \approx 0.1$
and $\gamma_4 \approx 0.01$ \cite{castroneto09}. We neglect spin-orbit
coupling and below we set $v_F=1$ for simplicity.

Hamiltonian \eqref{eq:HB} is invariant under time reversal symmetry (complex conjugation) combined with reflection from plane $p_y=0$ ($\phi \rightarrow -\phi$). This symmetry supports the $Z_2$ topology of Dirac points in graphene (Fig. \ref{Fig:BilayerGraphene} {\it right}) and Dirac lines in graphite.

\section{Point of merging of Dirac lines}

Since the  merging point $P^*$ is on the H-K-H line,  we consider the Hamiltonian at $p=0$ which has the following form (we consider only the traceless part):
\begin{eqnarray}
\tilde H(p=0,p_z) =H(p=0,p_z)  - \frac{1}{4} {\bf Tr} H(p=0,p_z) =
\nonumber
\\
  \begin{pmatrix} 
f(p_z) &0& 0& -2g(p_z)
\\ 
0& -f(p_z)& 0 &0
\\
0&  0  &-f(p_z) & 0
 \\
-2g(p_z)  & 0 &0 &f(p_z) 
\end{pmatrix} ~,
\nonumber
\\
 f(p_z)=\frac{1}{2}(\Delta - \gamma_2(p_z)),~~g(p_z)=\gamma_1\Gamma(p_z) \,.
\label{eq:tildeHB2}
\end{eqnarray}
This can be rewritten in terms of the Pauli matrices $\sigma_i$ and $\tau_i$:
\begin{equation}
\tilde H(p=0,p_z)=f(p_z) \tau_3\sigma_3 
+ g(p_z) (\tau_2\sigma_2 - \tau_1\sigma_1)  \,.
\label{eq:Pauli}
\end{equation}
Let us introduce new matrices $\tilde\tau_3=-\tau_3$,  $\tilde\tau_1=-\tau_1$,  $\tilde\tau_2=\tau_2$ and  the "total spin" ${\bf J}=\frac{1}{2}( {\mbox{\boldmath$\sigma$}} + \tilde{\mbox{\boldmath$\tau$}})$:
\begin{equation}
\tilde H(p=0,p_z)=- (f(p_z)+g(p_z)) \tilde\tau_3\sigma_3 
+ g(p_z)  {\mbox{\boldmath$\sigma$}} \cdot  \tilde{\mbox{\boldmath$\tau$}} \,,
\label{eq:Pauli2}
\end{equation}
\begin{equation}
 {\mbox{\boldmath$\sigma$}} \cdot  \tilde{\mbox{\boldmath$\tau$}}= 2J(J+1) - 3
~,~\tilde\tau_3\sigma_3  =  2J_3^2 -1\,.
\label{eq:J}
\end{equation}
The Hamiltonian at $p=0$ can be expressed in terms of quantum numbers $J$ and $J_3$:
\begin{equation}
\tilde H(p=0)=- (f(p_z)+g(p_z))(2J_3^2-1)
+ g(p_z)(2J(J+1) -3) \,.
\label{eq:HJ}
\end{equation}
The exceptional point $p_z=P^*$ is at $f(p_z)+g(p_z)=0$, where three
branches of the spectrum with $J=1$ are degenerate. Its position  is determined by equation
\begin{equation}
\Delta - \gamma_2(p_z)+2\gamma_1\Gamma(p_z) =0\,.
\label{eq:TerminationPoint}
\end{equation}
In other notations this is the known equation for the exceptional point in graphite, see e.g. Ref.~\cite{Mikitik2008}. It is situated in the vicinity of the H-point.
Away from this point, two branches with $J_3= \pm 1$ remain degenerate for all $p_z$.
Close to the exceptional  point 
\begin{equation}
\tilde H(p=0,J=1) \approx -\lambda(p_z-P^*)(2J_3^2-1)+  g(P^*)
 \,,
\label{eq:HJexpansion}
\end{equation}
where $\lambda=f'(P^*)+g'(P^*)$. 

The exceptional  point $P^*$ is the merging point of four Dirac lines
living in the symmetry planes at $p_z<P^*$. At $p_H >p_z>P^*$ the
spectrum remains degenerate: the branches with $J_3=+1$ and
$J_3=-1$ have the same energy. But these are the band contact lines rather than the Dirac
lines: the branches have the same sign of energy counted from the line
position, as distinct from $p_z<P^*$, where the touching branches have
energies with opposite sign, which is the characteristics of a Dirac
line (see Fig.~\ref{fig:finiteenergy}). This means that at   $p_z=P^*$ a band inversion occurs. The band inversion simultaneously happens for the branch with quantum numbers $J=1$ and $J_3=0$.

\section{2D Dirac fermions at H-point}
\label{DiracFermions}

To see that the topology at $p_H >p_z>P^*$ is trivial, let us consider
the $p_{x/y}$ plane containing the H-point. Since $\Gamma(p_H)=0$, the Hamiltonian at the plane $p_z=p_H$ is
\begin{equation}
\tilde H(p,H) = \frac{1}{2}(\Delta-\gamma_2(p_H))\tau_3\sigma_3 + p_x\sigma_1 + p_y \sigma_2
 \,.
\label{eq:Hpoint}
\end{equation}
It represents two copies of massive 2D fermions with Dirac spectrum:
\begin{equation}
\tilde E(p) =\pm \sqrt{M^2+p^2}  ~~,~~M=\frac{1}{2}(\Delta-\gamma_2(p_H))
 \,.
\label{eq:MassiveDirac}
\end{equation}
and thus two doubly degenerate branches. 
In addition to the contact line of $J_z=\pm 1$ branches, the other two branches, $J=1$, $J_3=0$  and  $J=J_3=0$, contact each other.
The Hamiltonian (\ref{eq:Hpoint}) has trivial $N_1$ topology. Due to continuity, the topology in the plane slightly below the H-point is also trivial. 
This demonstrates that at $p_H>p_z>P^*$ there are no Dirac lines. Instead there are band-contact lines, with trivial topological charge $N_1=0$. Thus the exceptional point $P^*$ is the point of merging of four Dirac lines.

\section{Finite-energy Dirac lines}
\begin{figure}[h]
\centering
\includegraphics[width=5.8cm]{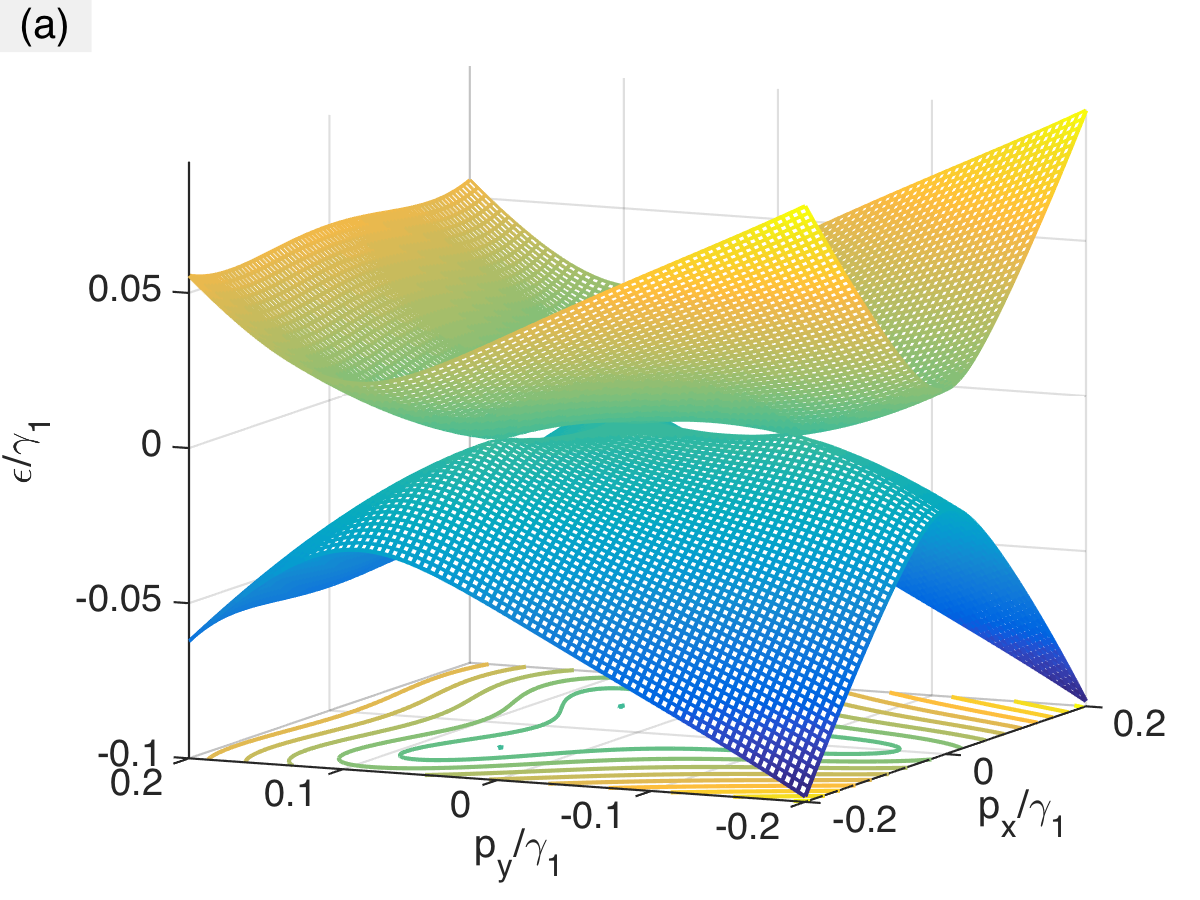}
\includegraphics[width=5.8cm]{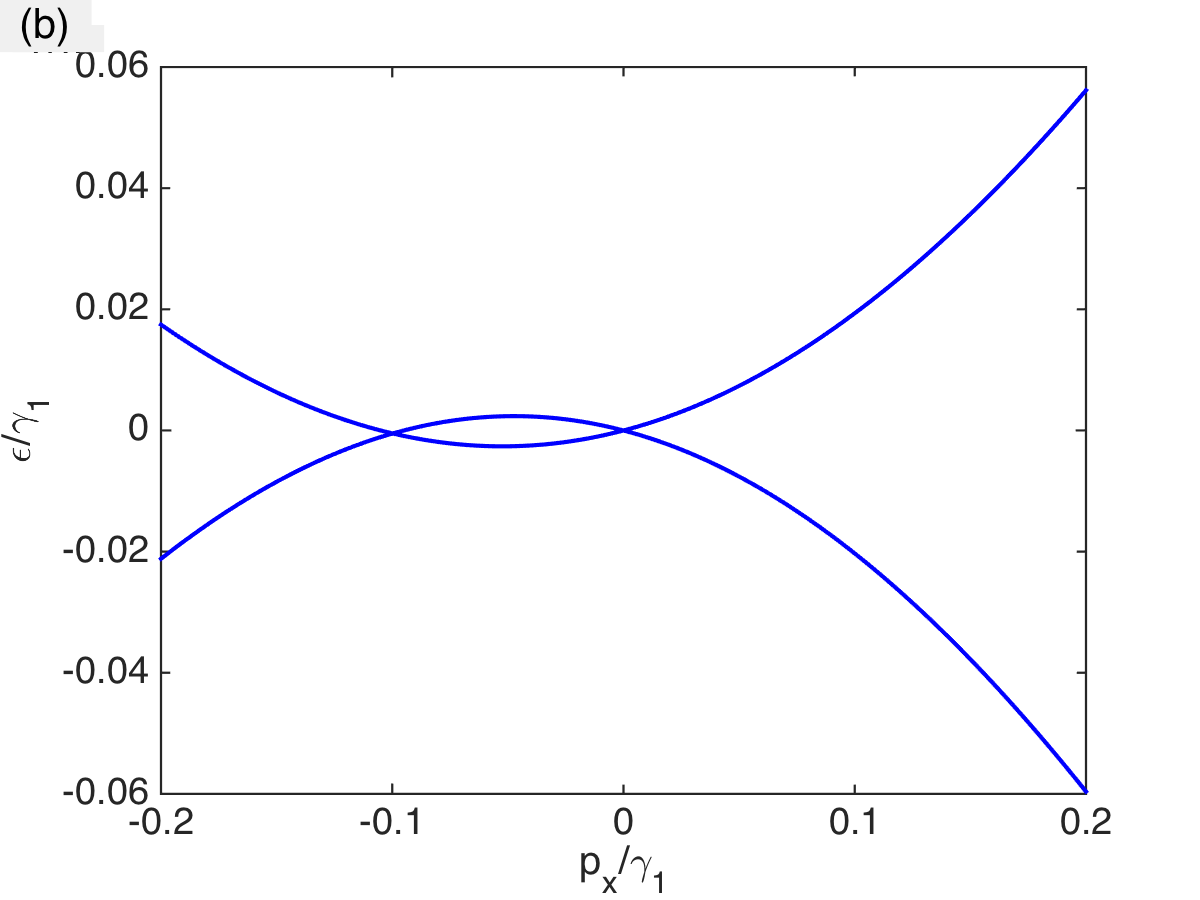}
\includegraphics[width=5.8cm]{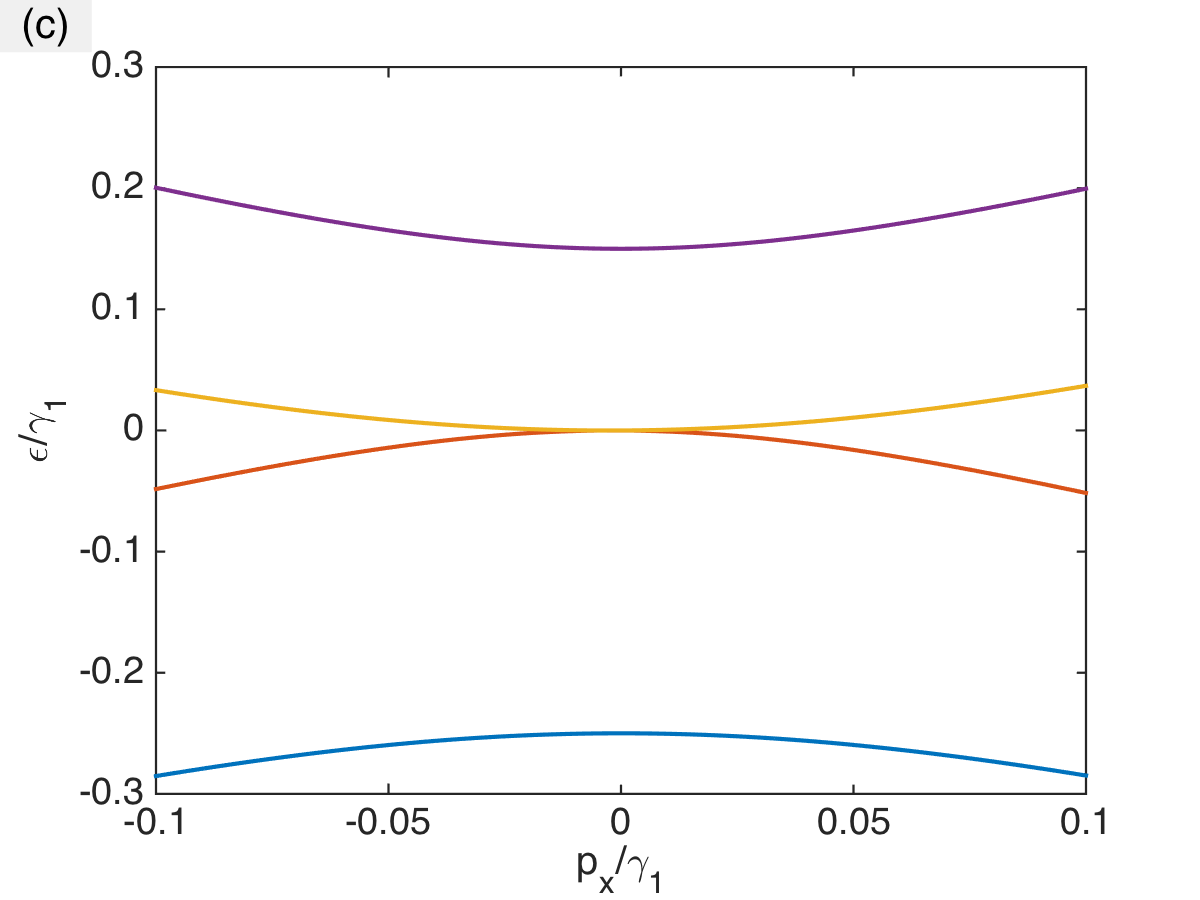}
\includegraphics[width=5.8cm]{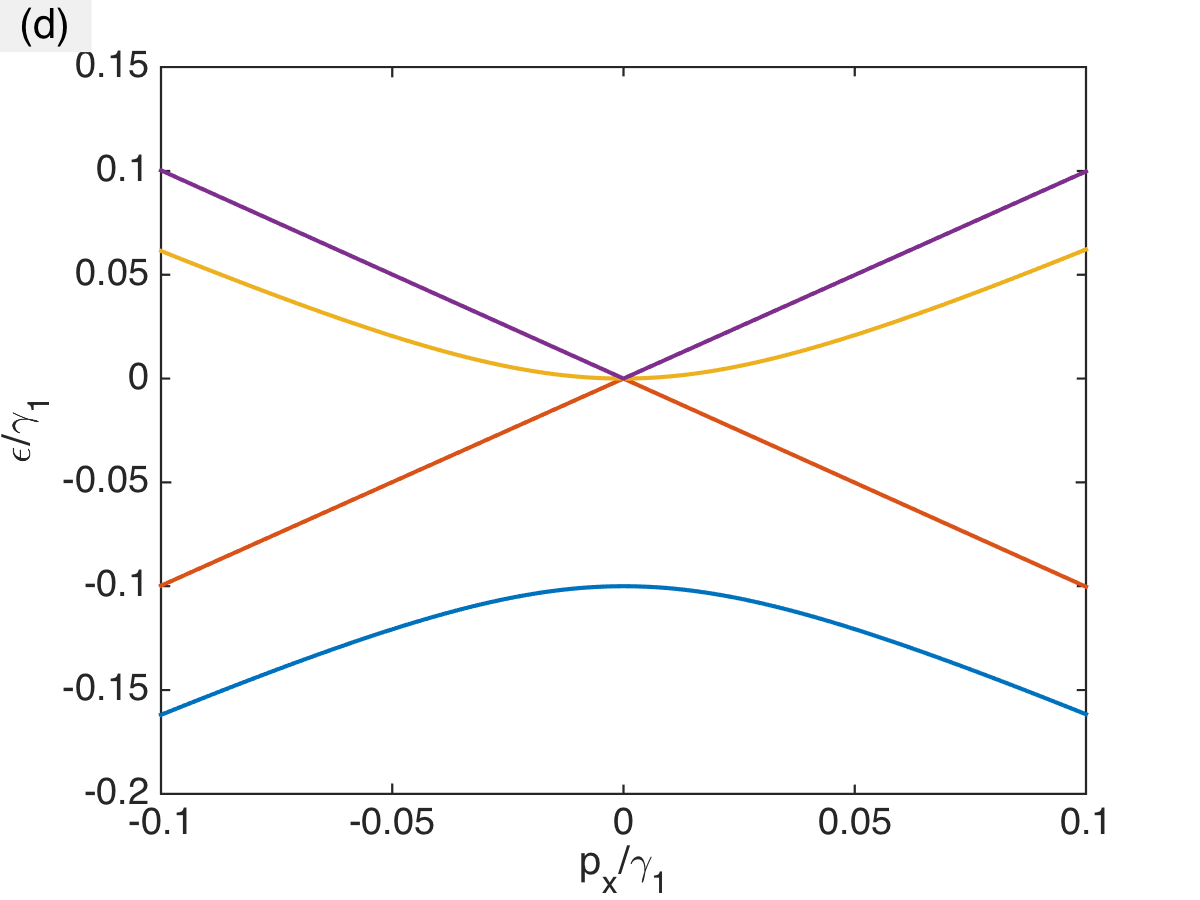}
\includegraphics[width=5.8cm]{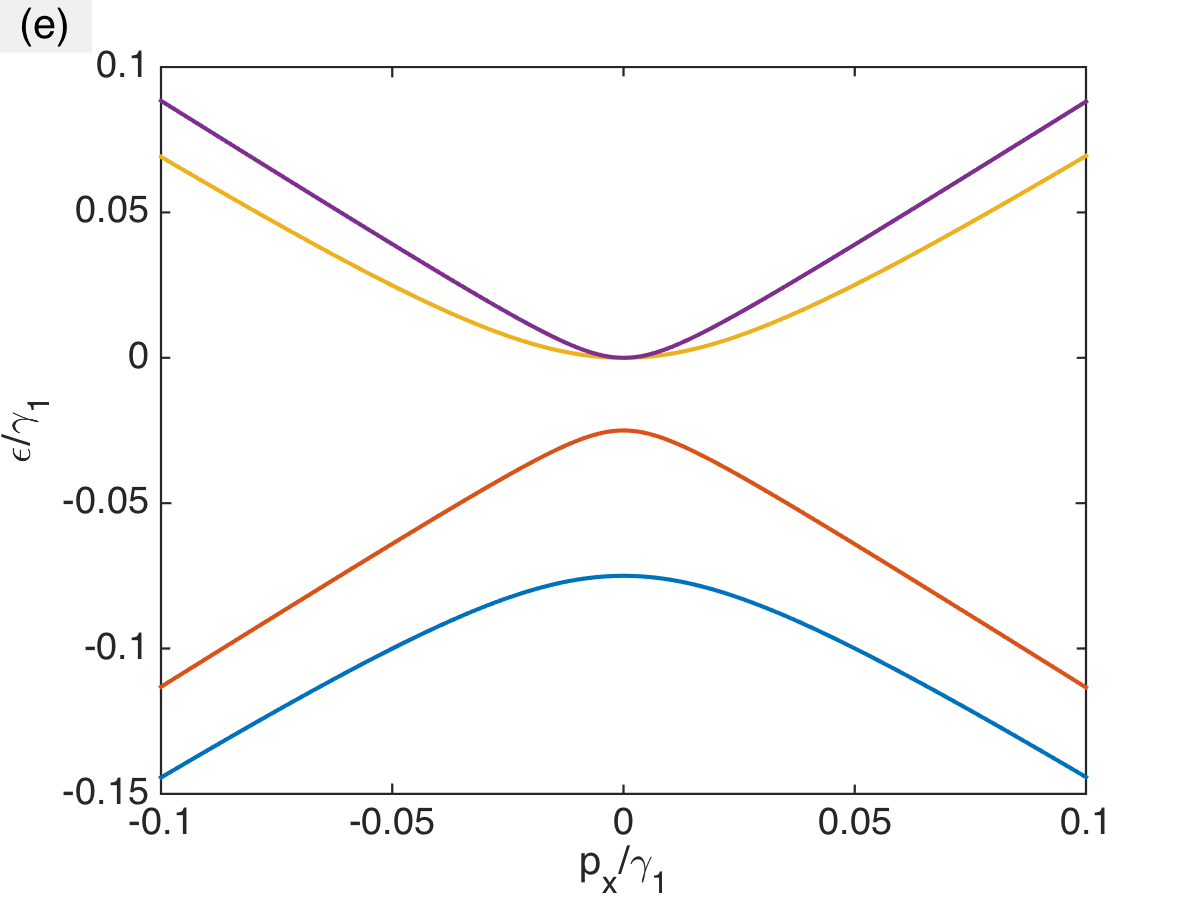}
\includegraphics[width=5.8cm]{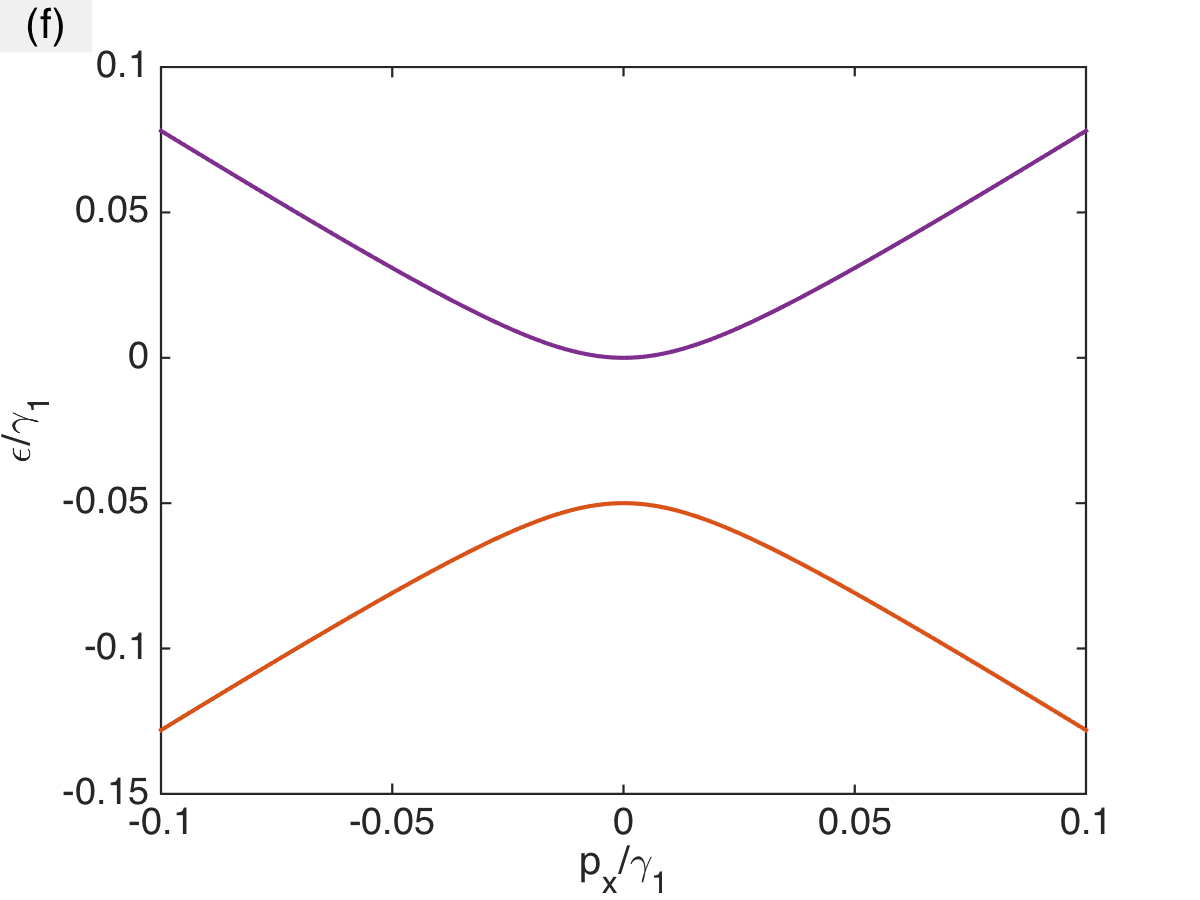}
\caption{(a) Positions of the Dirac lines in the energy spectrum are
  seen as the points in momentum space where the two eigenenergies
  meet (plotted for $\Gamma(p_z)=0.5$). (b) Cut through the
  $\phi=0$ axis revealing the positions of two of the
  Dirac lines. With a large $|\Delta|$ they shift towards lower values
  of $p$. (c) Same as previous but with $\Gamma(p_z)=-2\Delta/\gamma_1$,
  showing also the higher-energy branches. Now the finite-momentum
  Dirac lines are too close to the center point to be visible. (d) Dispersion at the
  nexus, $\Gamma(p_z)=-\Delta/(2\gamma_1)$; (e) above the
  nexus, $\Gamma(p_z)=-\Delta/(4 \gamma_1)$ and (f) at
  the $H$-point $\Gamma(p_z)=0$. The figures are computed with
  $\Delta=-0.05 \gamma_1$, $\gamma_3=0.1$, $\gamma_2=\gamma_4=0$. \label{fig:finiteenergy}
}
\end{figure}

Let us consider the occurrence of the Dirac lines at a finite
energy. For simplicity, we neglect $\gamma_2(p_z)$ and
$\gamma_4$, as they do not change the qualitative behavior of the
Dirac lines close to the nexus. In this case, the characteristic
equation for the bulk spectrum reads
\begin{equation}
(p^2+(\Delta-\epsilon) \epsilon)^2+(-4 p^3 \gamma_3^2
(\Delta-\epsilon)^2-4\gamma_1^2 \epsilon^2 + 8 p^3 \gamma_1 \gamma_3
\cos(3 \phi)) \Gamma^2 (p_z)+16 p^2 \gamma_1^2 \gamma_3^2
\Gamma(p_z)^4=0.
\end{equation}
This has four solutions for $\epsilon$. Let us analyze the solutions
for three cases: (a) $|p_z| \ll P^*$, (b) $|p_z|=P^*$, and (c),
$\pi/(2a) \ge |p_z| >
P^*$. The relevant eigensolutions for a specific set of parameters in the three
cases are plotted in Fig.~\ref{fig:finiteenergy}.

For region (a), there are two eigenenergies $\epsilon \approx \pm
\gamma_1 \Gamma(p_z)$, and two closer to zero. The latter two are plotted in
Figs.~\ref{fig:finiteenergy}(a,b) for a fixed $\Gamma(p_z)$. Two
low-energy solutions touch at specific points in the transverse
momentum direction, either at $p=0$ and close to $\cos(3\phi)=-1$,
$p=p_0=4\gamma_1 \gamma_3 \Gamma^2(p_z)$, which is the exact position of
the Dirac line for $\Delta = 0$. For small $\Delta$, the Dirac lines
at finite momentum obtain a finite energy $\epsilon=\Delta[1-1/(1+4
\gamma_3^2 \Delta \Gamma^2(p_z))]$ and slightly shift the line away
from $p=p_0$. The Dirac line character however persists. Figure
\ref{fig:finiteenergy}(c) shows all four eigenenergies in region
(a). Here the splitting of the two low-energy lines is not visible as
they are quite close to each other.

For $p_z=P^*$, the four solutions are plotted in Fig.~\ref{fig:finiteenergy}(d). For low $\Delta$ and $p$,
the energies are given by
\begin{equation}
\epsilon_{1,2}=\Delta \pm \sqrt{\Delta^2+p^2}, \epsilon_{3,4}=\pm p.
\end{equation}
There are hence two gapped and quadratic and two gapless linear
branches, and three of them meet at $\epsilon=p=0$. 

Above the nexus (region (c)), we may first set $\Gamma(p_z)=0$, i.e.,
consider the $H$ point. There, the
finite-energy solutions are given by (shifted by a constant energy from Eq.~\eqref{eq:MassiveDirac})
\begin{equation}
\epsilon_{a,b}=\frac{1}{2}(\Delta \pm \sqrt{4 p^2+\Delta^2}),
\end{equation}
The energies, plotted in Fig.~\ref{fig:finiteenergy}(f), are doubly degenerate, but the
degeneracy is lifted for a finite $\Gamma(p_z)$, yielding two pairs of
quadratic branches, the gap depending on the precise value of
$\Gamma(p_z)$. The lifting of this degeneracy is shown in Fig.~\ref{fig:finiteenergy}(e).

\section{Nexus}

In the real space, the nexus is a kind of a Dirac monopole that terminates cosmic strings (the so-called $Z$-string in the Standard Model).\cite{Cornwall1999} In chiral superfluid $^3$He-A the nexus  is a hedgehog which terminates vortices with the $Z_2$ or $Z_4$ topology (Fig. 17.3 in the book \cite{volovikbook} and Fig. \ref{Fig:Nexus} {\it top}).
In the dipole-locked $^3$He-A the homotopy group for vortices is $Z_2$. That is why two vortices can terminate at the hedgehog in the field of the orbital momentum vector $\hat{\bf l}$, which is the analog of the Dirac magnetic monopole.

In the same manner the Dirac lines can be described by the $Z_2$ group,
and if so, they may terminate on a Dirac monopole in some field. 
If  the invariant $N_1$ belongs to the $Z_2$ group, which means that $N_1=2$ and $N_1=0$ are equivalent, then by some transformation of hopping  elements the line described by even $N_1=2k$ can disappear. The lines with $N_1=\pm 1$ are stable. But when four such lines meet each other, their total charge is even, and thus they may annihilate each other. This also happens, since at $p_z>P^*$  the band contact lines are topologically trivial, $N_1=0$. 

So, the point $p_z=P^*$ represents the nexus in momentum space, see  Fig. \ref{Fig:Nexus} {\it bottom}.
It is distinct from the 3D Weyl  point in momentum space, which also
represents the momentum-space analog of a Dirac monopole. Such a monopole contains the Dirac string: a singularity in the Berry phase which terminates at the monopole, see Fig. 11.4 in Ref.~\cite{volovikbook}. But this string is not observable.

\section{Conclusion}
\label{conclusion}

The Dirac lines are suggested to exist in different 
semimetals. \cite{Weng2014,Xie2015,Kim2015,Yu2015,Uchoa2015,Fang2015,GosalbezMartinez2015}
The materials with Dirac lines are important, because
they may have an (approximate) flat band on the boundary or at the interface between  materials with different topological properties. The high density of states in the flat band provides a possible route to room-temperature superconductivity. \cite{HeikkilaVolovik2015} 

However, the Bernal graphite is a very instructive example.
Its spectrum demonstrates the possible interplay of several relevant topological invariants. One of them characterizes the local  line element as in Ref.~\cite{Horava2005}.  Another one is the global invariant, which characterizes for example the closed nodal ring as a whole: it is the same invariant which characterizes the 3D Dirac or Weyl point obtained by shrinking the nodal ring.  There are also the topological invariants that characterize the Fermi surface(s). These include the local invariant of the Fermi surface \cite{volovikbook} and the topology of the shape of the Fermi surface 
\cite{NovikovMaltsev1998}. All this may combine to produce exotic topological patterns. 
The interplay of Weyl point and Fermi surface topologies with exchange
of the Berry flux, when two Fermi surfaces contact each other, have been discussed in Refs.~\cite{KlinkhamerVolovik2005,GosalbezMartinez2015}.
The Bernal graphite provides an example of the interplay of Dirac
lines in the mirror planes, an exceptional point (nexus), and Fermi surfaces with touching points between electron and hole pockets,  see 
Refs.~\cite{Mikitik2006,Mikitik2008}. 
 
\section*{Acknowledgments}
This work was supported by the Academy of Finland through its Center
of Excellence program and by the European Research Council (Grant No. 240362-Heattronics).


\section*{References}

 
\begin{thebibliography}{99}


\bibitem{Neumann1929}
J. von Neumann  und E.P. Wigner,
\"Uber das Verhalten von Eigenwerten bei adiabatischen Prozessen,
 Phys. Zeit. {\bf 30}, 467--470 (1929).

\bibitem{Horava2005}  
P. Ho\v{r}ava,
Stability of Fermi surfaces and $K$-theory,
Phys. Rev. Lett. \textbf{95}, 016405 (2005).

\bibitem{Volovik2011} 
G.E. Volovik,
The topology of quantum vacuum, 
 Lecture Notes in Physics,  {\bf 870}, 343--383 (2013),
 arXiv:1111.4627.

\bibitem{Ryu2002}
S. Ryu and  Y. Hatsugai, 
Topological origin of zero-energy edge states in particle-hole symmetric systems,
 Phys. Rev. Lett. {\bf 89}, 077002 (2002).

\bibitem{SchnyderRyu2011}
 A.P.  Schnyder and S. Ryu, 
Topological phases and flat surface bands in superconductors without inversion symmetry,
Phys. Rev. B {\bf 84}, 060504(R) (2011).

\bibitem{SchnyderBrydon2015}
A.P. Schnyder and P.M.R. Brydon,
Topological surface states in nodal superconductors,
J. Phys.: Condens. Matter {\bf 27} 243201 (2015).

\bibitem{Volovik1993}
G.E. Volovik, 
Superconductivity with lines of gap nodes: Density of states in the vortex,
JETP Lett.  {\bf 58}, 469--473  (1993).

 \bibitem{HeikkilaVolovik2011} 
T.T. Heikkil\"a and G.E. Volovik,
Dimensional crossover in topological matter: Evolution of the multiple Dirac point in the layered system to the flat band on the surface,
Pis'ma ZhETF {\bf 93}, 63--68 (2011); JETP Lett. {\bf 93}, 59--65 (2011);
arXiv:1011.4185.

 \bibitem{HeikkilaKopninVolovik2011} 
T.T. Heikkil\"a, N.B. Kopnin and G.E. Volovik,
Flat bands in topological media, 
Pis'ma ZhETF {\bf 94}, 252-- 258 (2011); JETP Lett. {\bf 94}, 233--239(2011);
 arXiv:1012.0905.

 \bibitem{mcclure57} 
J.W. McClure, 
Band structure of graphite and de Haas-van Alphen effect, 
Phys. Rev. {\bf 108}, 612-618 (1957).

\bibitem{Pierucci2015} 
D. Pierucci, H. Sediri,  M. Hajlaoui, J.-C. Girard, T. Brumme, 
M. Calandra,  E. Velez-Fort,  G. Patriarche, M.G. Silly, G. Ferro,
V. Souliere, M. Marangolo, F. Sirotti,  F. Mauri,  and A. Ouerghi,
Evidence for flat bands near the Fermi level in epitaxial rhombohedral multilayer graphene,
ACS Nano {\bf 9}, 5432 (2015).

\bibitem{KlinkhamerVolovik2005a}
 F.R. Klinkhamer and G.E. Volovik, 
 Emergent CPT violation from the splitting of Fermi points, 
 Int. J.  Mod. Phys. A {\bf 20}, 2795--2812 (2005); 
hep-th/0403037.

\bibitem{CannFalko2006} 
E. McCann and V.I. Fal'ko,
Landau-level degeneracy and quantum Hall effect in a graphite bilayer,
Phys. Rev. Lett. {\bf 96}, 086805 (2006).

\bibitem{KoshinoAndo2006} 
M. Koshino and T. Ando,
Transport in bilayer graphene: Calculations within a self-consistent Born approximation,
Phys. Rev. B {\bf 73}, 245403 (2006).

 \bibitem{Mikitik2006} 
G.P. Mikitik and Yu.V. Sharlai, 
Band-contact lines in the electron energy spectrum of graphite,
Phys. Rev. B {\bf 73}, 235112 (2006).

\bibitem{Mikitik2008} 
G.P. Mikitik and Yu.V. Sharlai,
The Berry phase in graphene and graphite multilayers,
Low Temp. Phys. {\bf 34}, 794--780 (2008).


\bibitem{castroneto09}
A.H. Castro Neto, F. Guinea, N.M.R. Peres, K.S. Novoselov, and
A.K. Geim, The electronic properties of graphene, Rev. Mod. Phys. {\bf
  81}, 109 (2009).


\bibitem{volovikbook}
G.E. Volovik, 
{\em The Universe in a Helium Droplet} 
(Oxford University Press, Oxford, UK, 2003).

\bibitem{Cornwall1999}  
J.M. Cornwall,   
Center vortices, nexuses, and the Georgi-Glashow model, 
Phys. Rev.  D. {\bf 59}, 125015 (1999).

\bibitem{Weng2014}
H. Weng, Y. Liang, Q. Xu, Yu Rui, Z. Fang, Xi Dai, Y. Kawazoe,
Topological node-line semimetal in three dimensional graphene networks,
Phys. Rev. B {\bf 92}, 045108 (2015).

\bibitem{Xie2015}
L. S. Xie, L. M. Schoop, E. M. Seibel, Q. D. Gibson, W. Xie, and R. J. Cava, 
Potential ring of Dirac nodes in a new polymorph of Ca$_3$P$_2$, 
APL Mat. {\bf 3}, 083602 (2015).

\bibitem{Kim2015}
Youngkuk Kim,  B.J. Wieder,  C. L. Kane and A.M. Rappe,
Dirac line nodes in inversion symmetric crystals,
Phys. Rev. Lett. {\bf 115}, 036806 (2015).

\bibitem{Yu2015}
Yu Rui, H. Weng, Z. Fang, Xi Dai, and X. Hu, Topological node-line
semimetal and Dirac semimetal state in antiperovskite Cu$_3$PdN,
Phys. Rev. Lett. {\bf 115}, 036807 (2015).

\bibitem{Uchoa2015} 
K. Mullen, B. Uchoa, and D.T. Glatzhofer, Line of Dirac Nodes in
Hyperhoneycomb Lattices, Phys. Rev. Lett. {\bf 115}, 026403 (2015).

\bibitem{Fang2015}
C. Fang, Y. Chen, H-Y. Kee, and L. Fu, Topological nodal line
semimetals with and without spin-orbital coupling, arXiv:1506.03449.


\bibitem{GosalbezMartinez2015}
D. Gosalbez-Martinez, I. Souza, D. Vanderbilt,
Chiral degeneracies and Fermi-surface Chern numbers in bcc Fe,
arXiv:1505.07727


 \bibitem{HeikkilaVolovik2015} 
T.T. Heikkil\"a and G.E. Volovik,
Flat bands as a route to high-temperature superconductivity in graphite,
arXiv:1504.05824.

\bibitem{NovikovMaltsev1998}
S.P.Novikov and A.Ya.Maltsev,
Topological phenomena in normal metals,
Physics - Uspekhi, {\bf 41}, 231--239  (1998).

\bibitem{KlinkhamerVolovik2005}
F.R. Klinkhamer and G.E. Volovik,
Emergent CPT violation from the splitting of Fermi points
Int. J. Mod. Phys. A{\bf 20}, 2795--2812 (2005). 

\end{thebibliography}
\end{document}